\documentclass[twocolumn,prb,aps,epsf,showpacs,superscriptaddress]{revtex4-1}

\usepackage{graphicx}
\usepackage{dcolumn}
\usepackage{bm}

\newcommand{\I}{\mathrm{i}}
\newcommand{\figwidtha}{1}

\newcommand{\dxxyy}{$d_{x^2 - y^2}$}
\newcommand{\dwave}{$d$-wave}
\newcommand{\pwave}{$p$-wave}
\newcommand{\hata}{$\hat a$}
\newcommand{\hatb}{$\hat b$}
\newcommand{\tc}{$T_\mathrm{c}$}

\newcommand{\cecoin}{CeCoIn$_5$}
\newcommand{\bfcecoin}{\bf CeCoIn$_\mathbf{5}$}
\newcommand{\cecoinsn}{CeCoIn$_{5-x}$Sn$_x$}
\newcommand{\ceirin}{CeIrIn$_5$}
\newcommand{\cemin}{CeMIn$_5$}
\newcommand{\cetin}{CeTIn$_5$}
\newcommand{\ybco}[1]{YBa$_2$Cu$_3$O$_{#1}$}

\begin{document}

\title{Nodal quasiparticle dynamics in the heavy fermion superconductor \bfcecoin\\ revealed by precision microwave spectroscopy}

\author{C.~J.~S.~Truncik}
\author{W.~A.~Huttema}
\author{P.~J.~Turner}
\affiliation{Department of Physics, Simon Fraser University, Burnaby, BC, V5A~1S6, Canada}
\author{S.~\"Ozcan}
\affiliation{Cavendish Laboratory, Madingley Road, Cambridge, CB3 0HE, United Kingdom}
\author{N.~C.~Murphy}
\author{P.~R.~Carri\`ere}
\author{E.~Thewalt}
\author{K.~J.~Morse}
\author{A.~J.~Koenig}
\affiliation{Department of Physics, Simon Fraser University, Burnaby, BC, V5A~1S6, Canada}
\author{J.~L.~Sarrao}
\affiliation{Los Alamos National Laboratory, Los Alamos, New Mexico 87545, USA}  
\author{D.~M.~Broun}
\affiliation{Department of Physics, Simon Fraser University, Burnaby, BC, V5A~1S6, Canada}
 
\begin{abstract}

\cecoin\ is a heavy fermion superconductor with strong similarities to the high-\tc\ cuprates, including quasi-two-dimensionality, proximity to antiferromagnetism, and probable $d$-wave pairing arising from a non-Fermi-liquid normal state.  Experiments allowing detailed comparisons of their electronic properties are of particular interest, but in most cases are difficult to realize, due to their very different transition temperatures.  Here we use low temperature microwave spectroscopy to study the charge dynamics of the \cecoin\ superconducting state.  The similarities to cuprates, in particular to ultra-clean \ybco{y}, are striking: the frequency and temperature dependence of the quasiparticle conductivity are instantly recognizable, a consequence of rapid suppression of quasiparticle scattering below \tc; and penetration depth data, when properly treated, reveal a clean, linear temperature dependence of the quasiparticle contribution to superfluid density.  The measurements also expose key differences, including prominent multiband effects and a temperature-dependent renormalization of the quasiparticle mass. 

\end{abstract}

\maketitle{} 

In \cecoin, evidence for $d$-wave pairing comes predominantly from experiments that infer the presence and location of nodes in the superconducting energy gap.  This includes power laws in zero-field heat capacity\cite{Petrovic:2001p391,Movshovich:2001vn,Tanatar:2005p405} and thermal conductivity\cite{Movshovich:2001vn,Tanatar:2005p405}, and the field-angle dependence of heat capacity\cite{Aoki:2004gr,An:2010ef}, thermal conductivity\cite{Izawa:2001p769} and quantum oscillations in the superconducting state\cite{Settai:2001p770}.  These experiments are supported by evidence for spin-singlet pairing (decreasing Knight shift below \tc\cite{Curro:2001wg,Kohori:2001df} and Pauli-limited upper critical field\cite{Tayama:2002cv}) and by the nature of the spin-resonance peak\cite{Stock:2008fc}.  However, the emerging picture of \dxxyy\ pairing symmetry in \cecoin\ is complicated by observations on the isoelectronic homologue \ceirin, which suggest a hybrid order parameter with both line nodes and point nodes\cite{Shakeripour:2007p772}.

\begin{figure*}[ht]
\includegraphics*[width= 0.85 \textwidth]{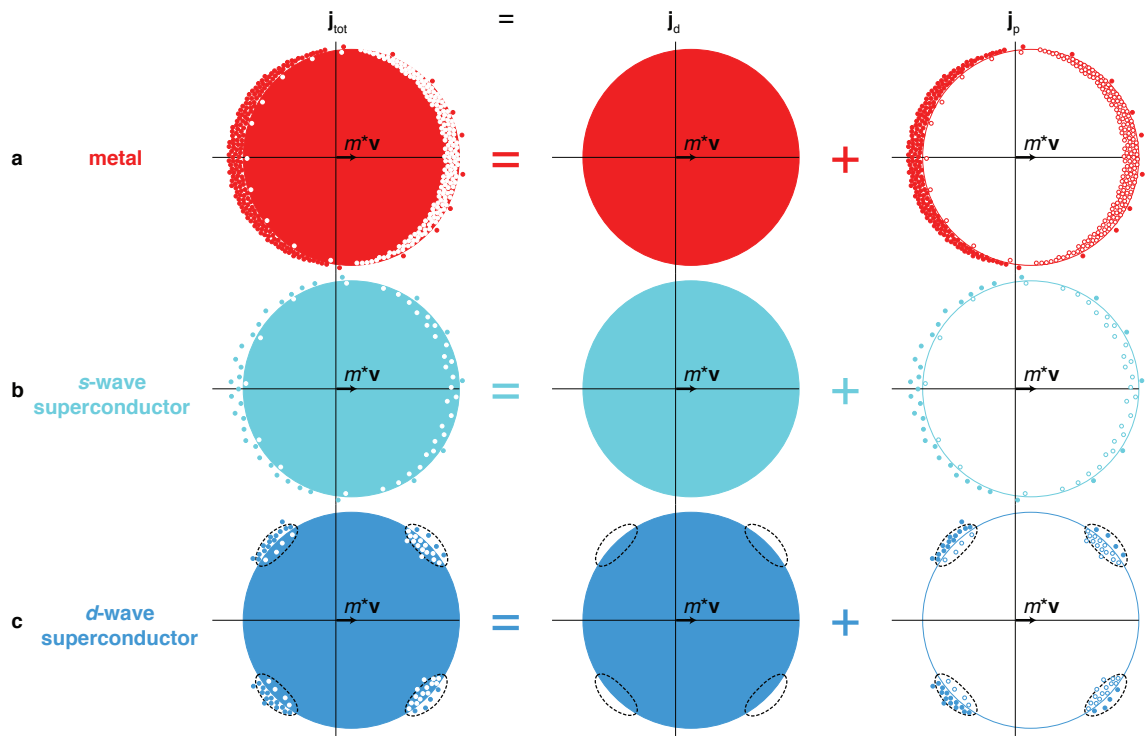}
\caption{\label{fig1} {\bf Superfluid density and the paramagnetic back-flow of excitations.} A momentum-space picture of the electron assembly in a superfluid density experiment, with electron-like excitations shown as filled circles and hole-like excitations as open circles.
Measurements of superfluid density probe the total nondecaying current in thermal equilibrium (left-hand figures), which can be decomposed into the sum of two terms\cite{Waldram:1996ta}.  Immediately after the application of a vector potential $\delta \mathbf{A}$, the Fermi sea is displaced by an amount $m^\ast \mathbf{v} =  - q \delta \mathbf{A}$ (central figures) giving rise to a diamagnetic current.  The vector potential also tilts the energy dispersion (\mbox{$\delta E_\mathbf{k} = - q \hbar/m^\ast \times \mathbf{k}\cdot \delta \mathbf{A}$}) and the resulting redistribution of electrons on the approach to equilibrium produces a paramagnetic back-flow current (right-hand figures).  ({\bf a}) The equilibrium current density in a metal is zero:  diamagnetic and paramagnetic currents are equal and opposite, and cancel.  ({\bf b}) The paramagnetic back-flow current in an $s$-wave superconductor is strongly suppressed by the opening of an isotropic energy gap.  The diamagnetic current is little changed from that in the normal state and a net current therefore flows in equilibrium, giving rise to a Meissner effect.  \mbox{({\bf c}) In a $d$-wave superconductor}, the paramagnetic back-flow consists predominantly of nodal quasiparticles, which make a linear-in-temperature contribution to superfluid density.  However, in \cecoin\ the diamagnetic contribution also has temperature dependence,  likely due to the material's proximity to a quantum critical point.  Isolating the nodal quasiparticle contribution requires that the current response be measured over a wide frequency range, with high frequencies probing the diamagnetic response and low frequencies the total current in equilibrium.  Measurements at intermediate microwave frequencies probe the transient processes that lead to the formation of the paramagnetic back-flow current, and provide a wealth of information on the charge dynamics of the nodal quasiparticles\cite{Bonn:2007hl}.}
\end{figure*}

\begin{figure*}[p]
\includegraphics*[width=0.95 \textwidth]{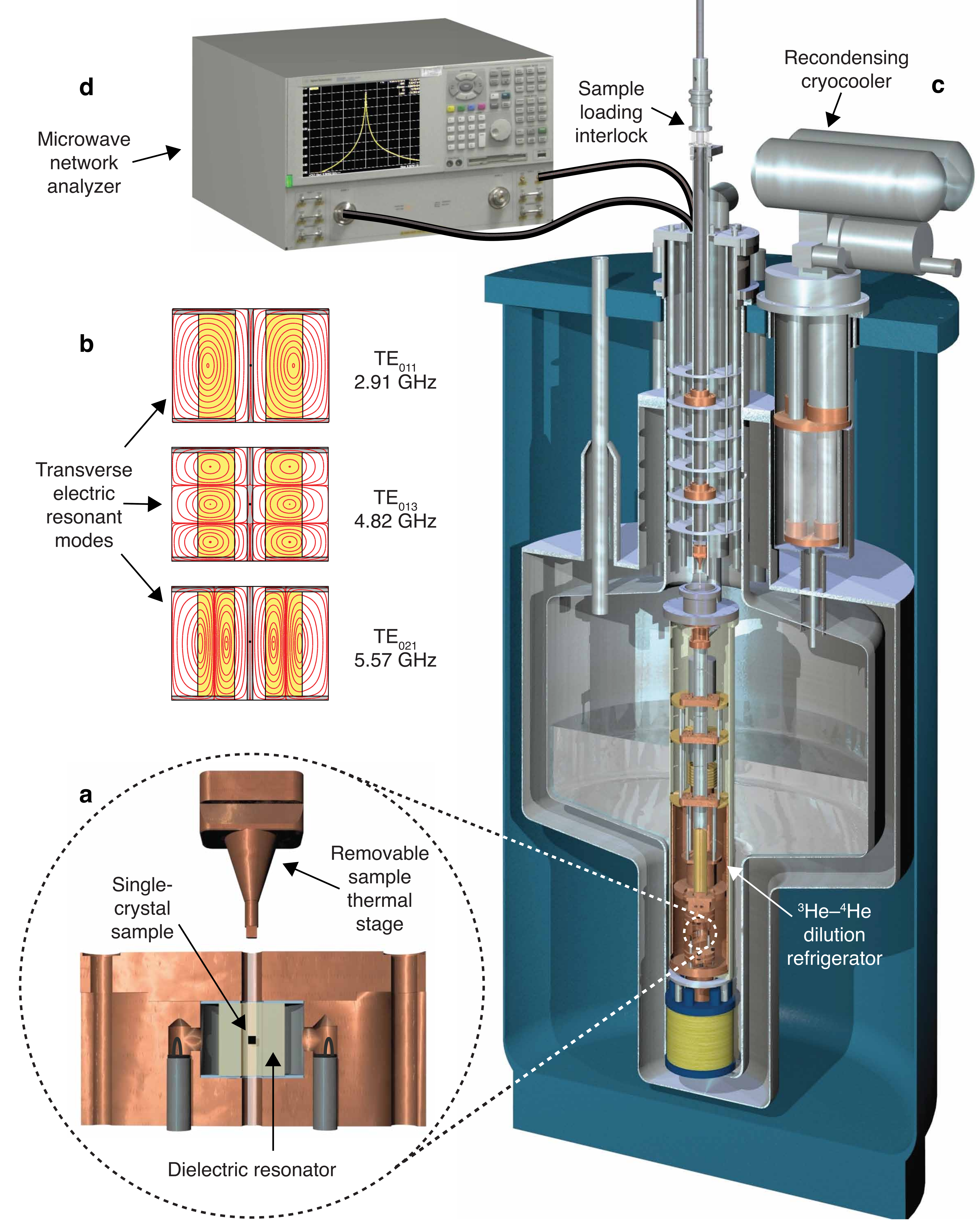}
\caption{\label{fig2} {\bf Millikelvin microwave spectroscopy.} ({\bf a}) A platelet single crystal of \cecoin\ is mounted on a removable thermal stage and introduced into a dielectric resonator.  ({\bf b}) The resonator is excited in multiple transverse electric (TE) modes at different frequencies, all characterized by a local maximum of the RF magnetic field (red lines) at the centre of the resonator.  This induces in-plane screening currents that flow across the broad faces of the \cecoin\ crystal.   ({\bf c})  The resonator is mounted below the mixing chamber of a $^3$He--$^4$He dilution refrigerator.  The sample stage is loaded from room temperature through a vacuum interlock, and can be cooled to 0.08~K.  A recondensing cryocooler eliminates helium boil off.  ({\bf d})  Shifts in sample surface impedance cause changes in resonance line shape that are read out by a low-noise microwave network analyzer.}
\end{figure*}

\begin{figure*}[ht]
\includegraphics*[width=\figwidtha \textwidth]{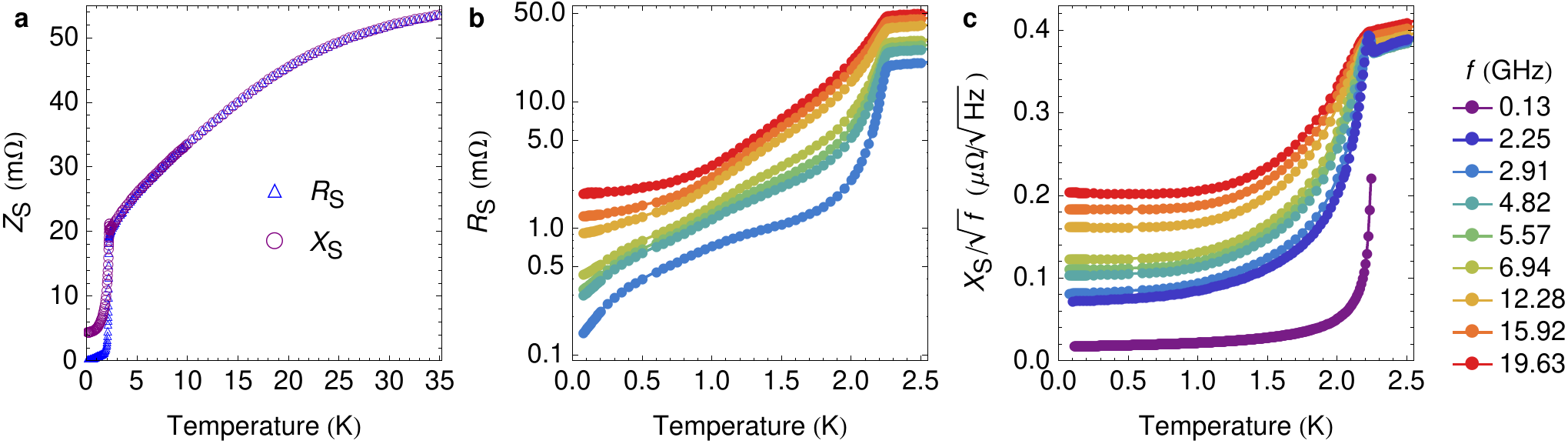}
\caption{\label{fig3} {\bf Microwave surface impedance.}   ({\bf a}) Surface impedance at 2.91~GHz, showing the results of the normal-state matching technique used to determine absolute reactance: $R_\mathrm{s}(T)$ and $X_\mathrm{s}(T)$ track well between $T = 10$~K and 35~K, a range in which the quasiparticle relaxation rate is much greater than the measurement frequency.  ({\bf b}) Surface resistance at frequencies from 2.91~GHz to 19.6~GHz, on a logarithmic scale. Absolute surface resistance is determined by a combination of cavity perturbation and \emph{in-situ}, resonator-based bolometry. ({\bf c}) Surface reactance, at all frequencies measured. For clarity, $X_\mathrm{s}(T)$ is scaled by $1/\sqrt{f}$, to factor out the expected frequency dependence well above \tc.}
\end{figure*}

Measurements of London penetration depth, $\lambda_\mathrm{L}$, should provide a particularly clean test of nodal structure, as $\lambda_\mathrm{L}$ is a thermodynamic probe that couples preferentially to itinerant electronic degrees of freedom\cite{Tinkham:1975un,Waldram:1996ta,Prozorov:2006wr,Bonn:2007hl}.  However, penetration depth data on \cecoin\ remain surprisingly unclear.  Instead of the linear temperature dependence expected for line nodes, all penetration depth measurements to date\cite{Ormeno:2002p404,Chia:2003et,Ozcan:2003p400} report temperature power laws ranging from $T^{1.2}$ to $T^{1.5}$.  This presents a conundrum --- mechanisms such as impurity pair-breaking\cite{PROHAMMER:1991p557,HIRSCHFELD:1993tf} and nonlocal electrodynamics\cite{Kosztin:1997p346} should cause a crossover to quadratic temperature dependence. Attempts to understand the behaviour in terms of impurity physics require unrealistically high levels of disorder\cite{Kogan:2009p399}. Here we solve this puzzle using comprehensive measurements of the frequency-dependent superfluid density.  These allow us to isolate the nodal-quasiparticle contribution to London penetration depth, revealing that it is accurately linear in temperature. 

To properly understand what microwave properties can tell us about a material\cite{Klein:1993p1129,Bonn:2007hl}, it is helpful to visualize the measurement process in the time domain.  The Meissner response of a superconductor is quantum mechanical in origin, and the fundamental electrodynamic relation is between the current density and the vector potential\cite{Waldram:1996ta}.  We therefore imagine a metal or superconductor perturbed by the sudden application of a vector potential $\delta \mathbf{A}$. As a result, all carriers experience an impulse $- q \delta \mathbf{A}$, where $q$ is the charge of the carriers.  The impulse sets the electron assembly into motion with average velocity $\mathbf{v} =  - q \delta \mathbf{A}/m^\ast$, where $m^\ast$ is the effective mass of the carriers.  This is sketched in the centre column of Fig.~\ref{fig1} and corresponds to a  current response that opposes the applied field. A measurement of current density  immediately after the application of the field reveals a diamagnetic contribution
\begin{equation}
\mathbf{j}_\mathrm{d} = - \frac{n q^2}{m^\ast} \delta \mathbf{A}\;,
\end{equation}
 where $n$ is the carrier density.  Note that the strength of the diamagnetic contribution is proportional to $n/m^\ast$ and is closely related to the plasma frequency of the carriers, $\omega_\mathrm{p} = \sqrt{nq^2/m^\ast \epsilon_0}$.  Here some care must be taken with the definition of `sudden':  if the carriers are excited with an arbitrarily sharp impulse, $n$ will be the total electron density, including core electrons, and $m^\ast$ the bare electron mass, devoid of interaction effects.  For practical purposes, a time scale is chosen that excites free carriers but avoids inter-band transitions.

In addition to inducing a diamagnetic current, $\delta \mathbf{A}$ changes the energy of the electron states, tilting the energy dispersion in $k$-space by an amount \mbox{$\delta E_\mathbf{k} = - q \hbar/m^\ast \times \mathbf{k}\cdot \delta \mathbf{A}$}. Immediately after excitation the electron assembly is therefore in a nonequilibrium configuration. Equilibrium is subsequently restored by scattering processes that transfer momentum to the crystal lattice. By studying the current response in this regime, we learn a great deal about electronic relaxation mechanisms in the material.

At times long enough for the electron assembly to have returned to equilibrium, it is useful to define the total current density as the sum of diamagnetic and paramagnetic pieces\cite{Waldram:1996ta},
$\mathbf{j}_\mathrm{tot} = \mathbf{j}_\mathrm{d} + \mathbf{j}_\mathrm{p}$. As discussed above, $\mathbf{j}_\mathrm{d}$ represents the instantaneous diamagnetic response intrinsic to the electronic states, with the tacit understanding that $\mathbf{j}_\mathrm{d}$ has been measured slowly enough to include only free carriers.  The paramagnetic part, $\mathbf{j}_\mathrm{p}$, is of a very different character --- it captures the change in current resulting from the reorganization of electrons in the new equilibrium state.  $\mathbf{j}_\mathrm{p}$ is therefore very sensitive to the details of the electronic energy spectrum, making it a powerful probe of pairing symmetry in a superconductor.  The way this process plays out is illustrated in Fig.~\ref{fig1}, for a metal; and for $s$-wave and $d$-wave superconductors.  For the metal in equilibrium, $\mathbf{j}_\mathrm{tot} = 0$: the paramagnetic redistribution of electronic occupation results in a back-flow current of equal but opposite magnitude to the initial diamagnetic shift.  For a superconductor, in contrast, the equilibrium current density in the presence of a magnetic field is non-zero --- there is a Meissner effect. 
The strength of the diamagnetic contribution is unchanged by the onset of superconductivity.  Instead, the opening of a superconducting gap dramatically weakens the paramagnetic response.
The paramagnetic term is strongly temperature dependent, in principle going to zero in a clean superconductor at zero temperature.  The form of this temperature dependence is highly sensitive to the structure of the energy gap, in particular to the presence of gap nodes.  For \cecoin, which is thought to be a \dwave\ superconductor with line nodes in the energy gap, the expected behaviour is a linear temperature dependence of $\mathbf{j}_\mathrm{p}$.  However, a complication now arises: the experimentally accessible quantity in a penetration-depth experiment is not the paramagnetic current density $\mathbf{j}_\mathrm{p}$, but the total current density $\mathbf{j}_\mathrm{tot}$.  Most experiments skirt this issue by assuming that the diamagnetic response, $\mathbf{j}_\mathrm{d}$, is temperature independent: it is difficult to measure directly, and in most superconductors has little temperature dependence anyway.  However, we will show below that this fundamental assumption breaks down in \cecoin, and is the reason why anomalous temperature power laws have been reported in penetration depth.  Our measurements reveal that the diamagnetic response (the plasma frequency) of \cecoin\  weakens on cooling, in a manner corresponding to an increase in quasiparticle effective mass.  That this occurs in \cecoin\ is not too surprising, as it is suggestive of proximity to a quantum critical point\cite{Sidorov:2002cr,Paglione:2006p2791}.

Although a time-domain picture provides a useful means of understanding electrodynamic measurements, the experiments themselves are usually carried out in the frequency domain, in our case using a set of discrete frequencies ranging from $\omega/2 \pi = 0.13$~GHz to 19.6~GHz.  
Low frequencies measure the long-time behaviour, and are sensitive to the equilibrium supercurrent density. High frequencies probe the short-time behaviour and, if carried out in a regime in which $\omega$ is greater than the electronic relaxation rate $1/\tau$, probe the instantaneous diamagnetic response and therefore the plasma frequency of the entire electron assembly.  At intermediate frequencies, much information can be obtained on the scattering dynamics of the thermally excited quasiparticles\cite{Bonn:2007hl}.  This is of particular interest in \cecoin\ because normal-state transport measurements reveal strong inelastic scattering and non-Fermi-liquid behaviour\cite{Kim:2001gd,Sidorov:2002cr}.   In the cuprates, where similarly strong scattering is observed in the normal state\cite{VARMA:1989p468}, electrodynamic measurements show a rapid collapse in quasiparticle scattering on cooling through \tc\cite{Nuss:1991wj,Bonn:1992fx,Hosseini:1999p383}, indicating that the charge carriers couple to a spectrum of fluctuations of electronic origin, in contrast to the phonons of a conventional metal.  Early measurements on \cecoin\  are suggestive of similar behaviour\cite{Movshovich:2001vn,Ormeno:2002p404}.\\

\begin{figure*}[ht]
\includegraphics*[width=\figwidtha \textwidth]{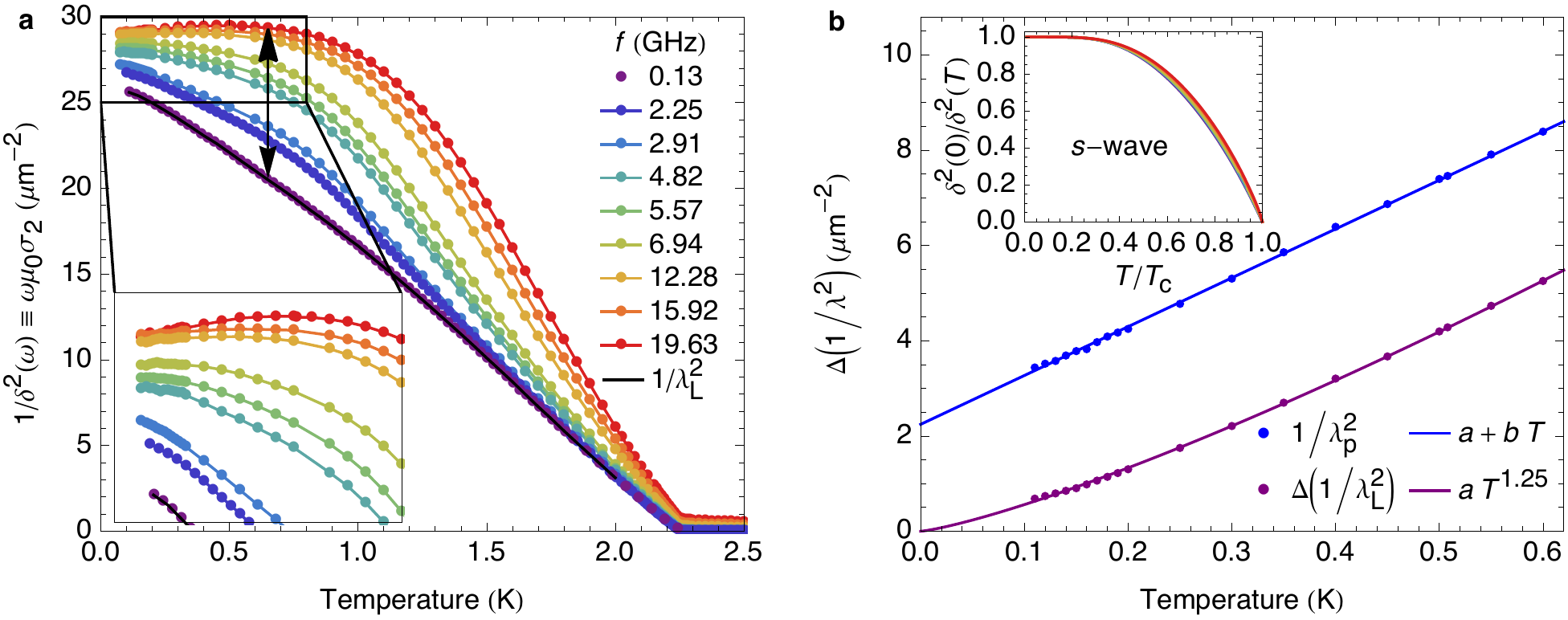}
\caption{\label{fig4} {\bf Superfluid density of \bfcecoin.}  ({\bf a}) 
Frequency-dependent superfluid density, $1/\delta^2(\omega,T) \equiv \omega \mu_0 \sigma_2(\omega,T)$, plotted as a function of temperature, for frequencies from 0.13 to 19.6~GHz.  $1/\lambda^2_\mathrm{L}(T)$, the zero-frequency limit of $1/\delta^2(\omega,T)$, is obtained from fits to complex conductivity spectra and lies on top of the 0.13~GHz data.  $\lambda_\mathrm{L}(T \to 0) = 1960$~\AA. The inset shows a close-up of the low temperature region, in which the temperature slope of $1/\delta^2(\omega,T)$ changes sign with increasing frequency. ({\bf b})  The  temperature-dependent part of the total superfluid density, $\Delta(1/\lambda^2_\mathrm{L}) \equiv 1/\lambda^2_\mathrm{L}(T \to 0) - 1/\lambda^2_\mathrm{L}(T)$, follows a $T^{1.25}$ power law.  The paramagnetic part of the superfluid density, $1/\lambda^2_\mathrm{p} \equiv 1/\delta^2(19.6\mbox{ GHz},T) - 1/\lambda^2_\mathrm{L}(T)$, isolates the contribution from nodal quasiparticles and follows a linear temperature dependence.  Its zero-temperature intercept indicates a residual, uncondensed spectral weight of 7\%.   Inset: the normalized superfluid density of an $s$-wave superconductor, calculated using Mattis--Bardeen theory\cite{MATTIS:1958p326} for the same set of reduced frequencies (and same colour scheme) as the \cecoin\ experiment.  In the $s$-wave case, the isotropic energy gap leads to exponentially activated behaviour at low temperatures.}
\end{figure*}

\noindent \textsf{\textbf{Results}}

\noindent {\bf Surface impedance.} Measurements of surface impedance, $Z_\mathrm{s} = R_\mathrm{s} + \I X_\mathrm{s}$, have been made using resonator perturbation\cite{1946Natur.158..234P,Altshuler:1963wq,Klein:1993p1129,Donovan:1993p1114,DRESSEL:1993p341,Huttema:2006p344,Bonn:2007hl}.  The sample, a small single crystal of \cecoin, is placed inside a dielectric resonator at a maximum in the RF magnetic field, as shown in Fig.~\ref{fig2}.  Screening currents are induced to flow near the sample surface and penetrate a skin depth $\delta$.  In the penetrated region, energy is stored both as field energy and as the kinetic energy of the superelectrons\cite{Bonn:2007hl} --- this leads to a surface reactance  $X_\mathrm{s} \approx \omega \mu_0 \delta$.  Field penetration changes the effective volume of the resonator and hence its resonant frequency\cite{Bonn:2007hl}.  Although superconductors have perfect DC conductivity, the finite inertia of the electrons means that accelerating them at high frequencies requires significant electric field at the sample surface: the strength of the field is determined by Faraday's law and grows in proportion to both the frequency $\omega$ and the depth of field penetration.  The electric field couples to quasiparticle excitations in the superconductor\cite{Bonn:2007hl} producing a surface resistance, $R_\mathrm{s}$, proportional to the power absorption.  This grows as the square of electric field, and therefore approximately as $\omega^2$.  This dissipation is measured by monitoring the quality factor of the resonator\cite{Huttema:2006p344,Bonn:2007hl,Klein:1993p1129}.  The complete set of surface impedance data is presented in Fig.~\ref{fig3}.\\

\noindent {\bf Microwave conductivity.} In the frequency domain, current density $\mathbf{j}$ is related to electric field $\mathbf{E}$ by a complex-valued microwave conductivity\cite{Klein:1993p1129,Bonn:2007hl}: \mbox{$\mathbf{j}(\omega) = \sigma(\omega) \mathbf{E}(\omega)$}.  In a superconductor, the dominant contribution to the complex conductivity is a purely imaginary response associated with the superfluid density: $\sigma_\mathrm{s} = 1/\I \omega \mu_0 \lambda_\mathrm{L}^2$.  There is an additional contribution to the conductivity, $\sigma_\mathrm{qp}$, arising from the non-equilibrium response of the quasiparticles as they relax back to equilibrium --- this derives from the transient response to the applied field and contains important information on relaxation mechanisms.  $\sigma_\mathrm{qp}$ is in general complex, but is predominantly real for low frequencies, $\omega \ll 1/\tau$, where it represents microwave power absorption, becoming imaginary at high frequencies, $\omega \gg 1/\tau$, where the field-screening effect of the quasiparticles becomes indistinguishable from that of the superfluid.  This leads to a two-fluid model of the microwave conductivity\cite{Bonn:2007hl}
\begin{equation}
\sigma(\omega,T) \equiv \sigma_1 - \I \sigma_2 = \frac{1}{\I \omega \mu_0 \lambda_\mathrm{L}^2(T)} + \sigma_\mathrm{qp}(\omega,T)\;.
\label{Eq_complex_conductivity}
\end{equation}
In our experiments,  the microwave conductivity  is obtained from the surface impedance assuming the local electrodynamic relation\cite{Klein:1993p1129,Bonn:2007hl} $\sigma = \I \omega \mu_0/Z_\mathrm{s}^2$.\\

\begin{figure*}[ht]
\includegraphics*[width=\figwidtha \textwidth]{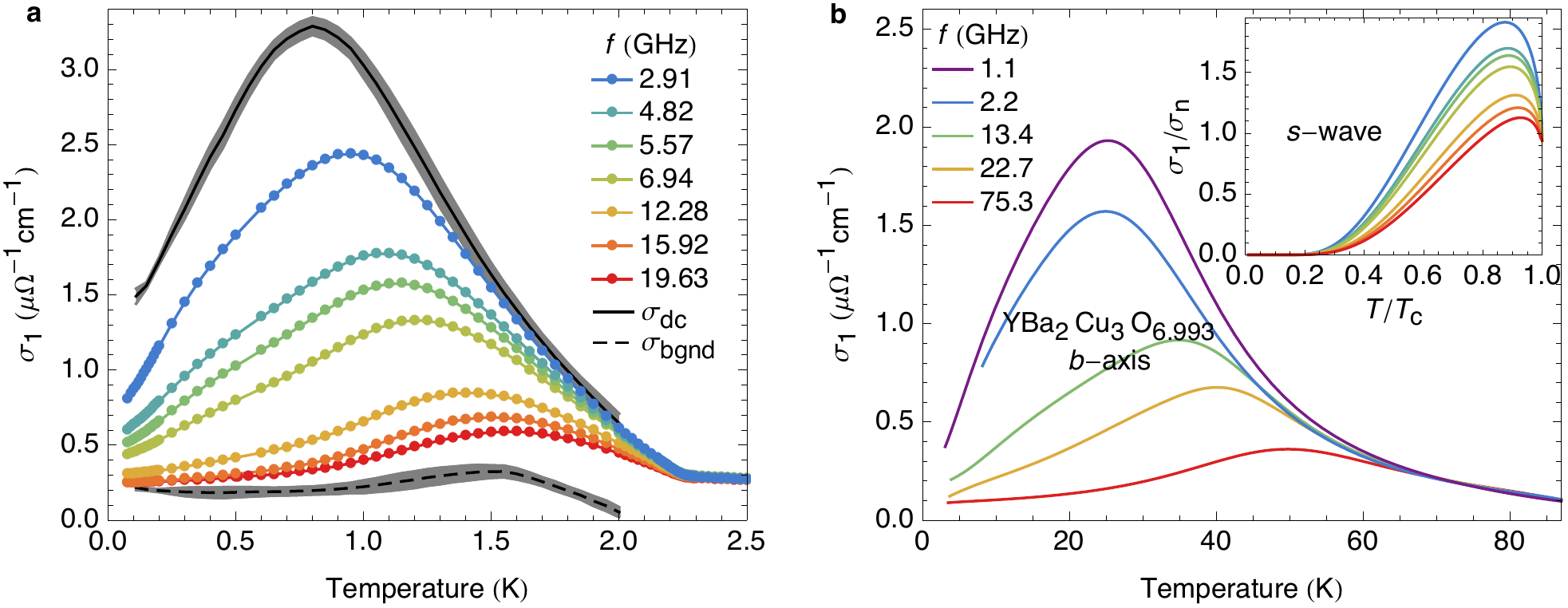}
\caption{\label{fig5} {\bf Quasiparticle conductivity of CeCoIn$_\mathbf{5}$ and YBa$_\mathbf{2}$Cu$_\mathbf{3}$O$_\mathbf{6.993}$.}  ({\bf a}) Real part of the conductivity, $\sigma_1$, of \cecoin\ as a function of temperature, for discrete frequencies from 2.91 to 19.63~GHz.  Also plotted are parameters from fitting to conductivity spectra: $\sigma_\mathrm{bgnd}(T)$; and $\sigma_\mathrm{dc}(T) = \sigma_0(T) + \sigma_\mathrm{bgnd}(T)$. Shaded confidence bands denote standard errors in these parameters. ({\bf b}) For comparison, the real part of the $b$-axis conductivity of $T_\mathrm{c} = 89$~K \ybco{6.993}, as a function of temperature, at frequencies from 1.1 to 75.3~GHz (data from Ref.~\onlinecite{Harris:2006p388}).  Inset: the normalized quasiparticle conductivity of an $s$-wave superconductor, calculated using Mattis--Bardeen theory\cite{MATTIS:1958p326} for the same set of reduced frequencies (and same colour scheme) as the \cecoin\ experiment: a prominent BCS coherence peak is observed immediately below \tc, with exponential freeze-out at low temperatures.}
\end{figure*}

\begin{figure*}[t]
\includegraphics*[width= \textwidth]{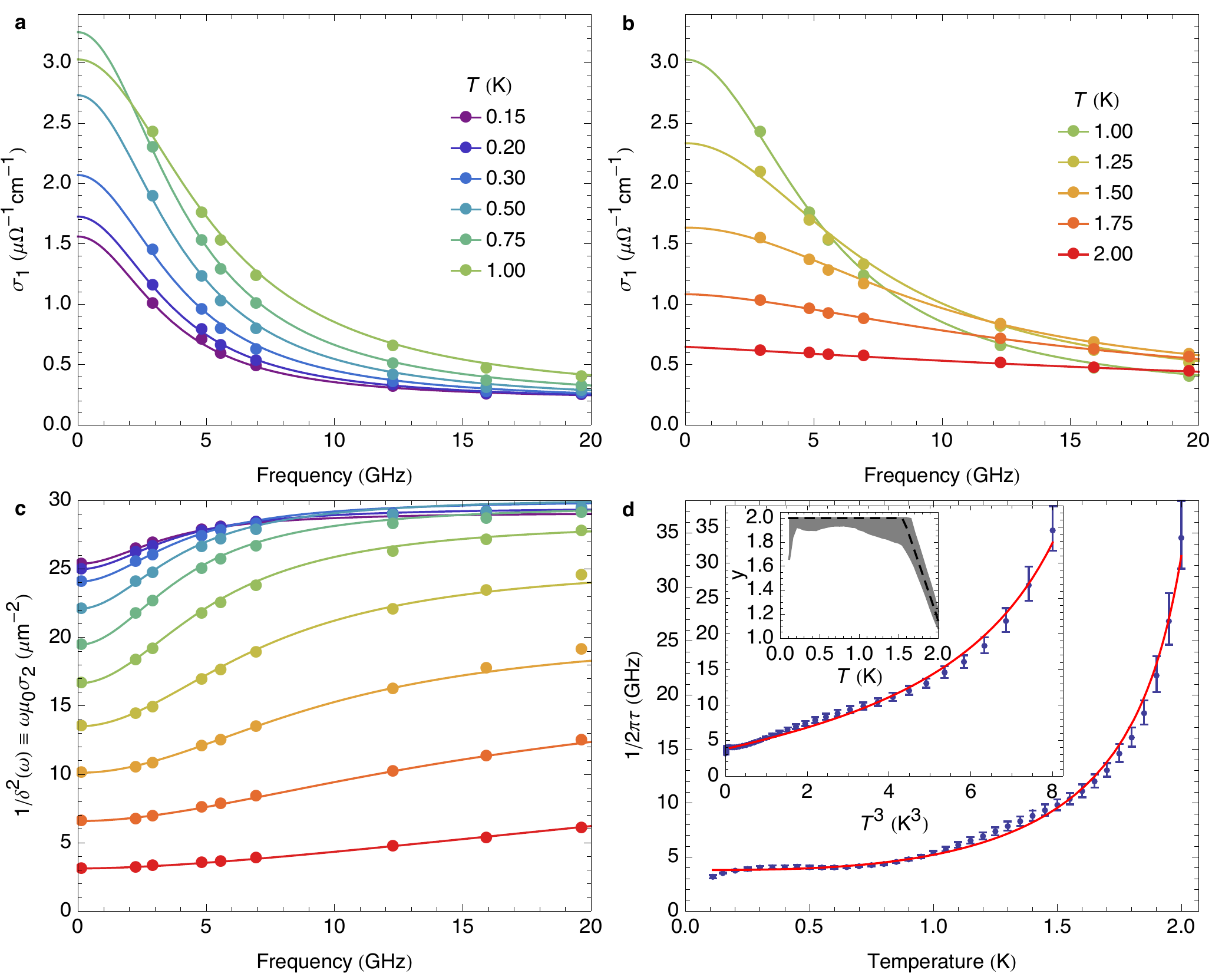}
\caption{\label{fig6} {\bf Complex conductivity spectra and model fits.} ({\bf a}),({\bf b}) Real part of the conductivity as a function of frequency, at discrete temperatures.  ({\bf c}) Frequency-dependent superfluid density at the same set of temperatures as in {\bf a} and {\bf b}.  The curves in {\bf a, b} and {\bf c} denote simultaneous fits to $\sigma_1(\omega)$ and $1/\delta^2(\omega)$ at each temperature, using the three-component conductivity model described in Methods.  ({\bf d}) The relaxation rate $1/\tau(T)$ obtained from the fits, as a function of temperature. The primary inset shows $1/\tau(T)$ as a function of $T^3$.  Vertical bars indicate standard errors.  The red lines are a fit to the function $\hbar/\tau(T) = \hbar/\tau_0 + A k_\mathrm{B}^3 T^3/\Delta^2(T)$, with $\Delta(T) = \Delta_0 \tanh\big(2.4 \sqrt{T_\mathrm{c}/T -1}\big)$, $\Delta_0 = 3~k_\mathrm{B} T_\mathrm{c}$ and $A = 3.36$.  The secondary inset shows $y(T)$, the best-fit frequency exponent of the modified Drude spectrum in  Eq.~\ref{Eq_conductivity}.  The shaded band denotes the 1-$\sigma$ confidence interval.  $y(T)$ is constrained to lie in the range  $1 < y \le 2$. }
\end{figure*}

\begin{figure*}[ht]
\includegraphics*[width=\figwidtha \textwidth]{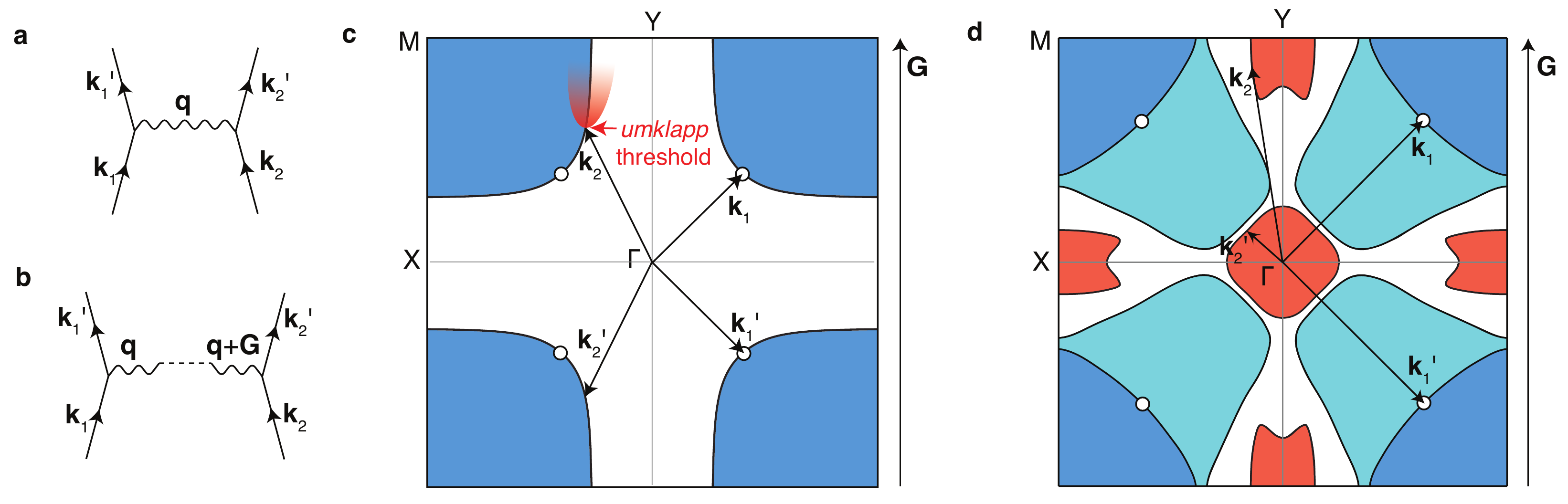}
\caption{\label{fig7} {\bf Quasiparticle--quasiparticle scattering and the \emph{umklapp} gap.}  ({\bf a}) Two quasiparticles, of wave vectors $\mathbf{k}_1$ and $\mathbf{k}_2$, interact and scatter into states $\mathbf{k}_1^\prime$ and $\mathbf{k}_2^\prime$, exchanging momentum $\mathbf{q}$.  Such a scattering process leaves the net momentum in the electron system unchanged ($\mathbf{k}_1 + \mathbf{k}_2 = \mathbf{k}_1^\prime + \mathbf{k}_2^\prime$) and is therefore ineffective at relaxing an electrical current.  ({\bf b}) In a solid, crystal momentum is only conserved to within a reciprocal lattice vector $\mathbf{G}$: i.e., $\mathbf{k}_1 + \mathbf{k}_2 = \mathbf{k}_1^\prime + \mathbf{k}_2^\prime + \mathbf{G}$.  \emph{Umklapp} processes, for which $\mathbf{G} \ne 0$,  transfer momentum from the electron assembly to the crystal lattice, and are very effective at relaxing an electrical current.  ({\bf c})  An \emph{umklapp} process in a single-band \dwave\ superconductor\cite{Walker:2000ux}.  A hole-like Fermi sea, characteristic of a cuprate superconductor, is shown shaded in blue, with nodes (open circles) on the zone diagonals.  In order to conserve crystal momentum, a near-nodal quasiparticle, $\mathbf{k}_1$, must be partnered with a second quasiparticle, $\mathbf{k}_2$, located well away from a node.  The energy threshold for this process is the ``\emph{umklapp} gap", and strongly suppresses quasiparticle--quasiparticle scattering at low temperatures.    ({\bf d}) The Fermi surface of \cecoin\ in the $k_z = 0$ plane\cite{XiaoWen:2011kg,Settai:2001p770}, illustrating the material's multiband nature.  An inter-band \emph{umklapp} process is shown, in which a nodal quasiparticle in a quasi-2D band (dark blue) scatters from an electron in one of the light, 3D Fermi pockets (red).  Thermodynamic experiments indicate that superconductivity in the light, 3D pockets is weak\cite{Tanatar:2005p405} and this should substantially reduce the \emph{umklapp} threshold.}
\end{figure*}

\noindent {\bf Superfluid density.} The static superfluid density, $1/\lambda^2_\mathrm{L}$, is obtained from the complex conductivity in the zero-frequency limit:
\begin{equation}
1/\lambda^2_\mathrm{L} = \lim_{\omega \to 0} \omega \mu_0 \sigma_2\;.
\end{equation}
At finite frequencies, $\mathrm{Im}\left\{\sigma_\mathrm{qp}(\omega,T) \right\}$ makes a significant contribution to $\sigma_2$: this screening effect is particularly prominent in \cecoin\ due to its long quasiparticle lifetimes.  In order to separate these two components, we define a frequency-dependent superfluid density,
\begin{equation}
\frac{1}{\delta^2(\omega,T)}  \equiv \omega \mu_0 \sigma_2(\omega,T) = \frac{1}{\lambda^2_\mathrm{L}(T)} - \omega \mu_0 \mathrm{Im}\left\{\sigma_\mathrm{qp}(\omega,T) \right\}.
\label{Eq_freq_dep_superfluid}
\end{equation}
The quasiparticle-relaxation contribution to $1/\delta^2(\omega)$ vanishes in the static limit, but is nonzero at all finite frequencies.  At high frequencies, \mbox{$\omega \mathrm{Im}\left\{\sigma_\mathrm{qp} \right\} \to n_\mathrm{qp} e^2/m^\ast$}, providing a measure of the quasiparticle density, $n_\mathrm{qp}$, and therefore the uncondensed oscillator strength in the quasiparticle spectrum.  Equation~\ref{Eq_freq_dep_superfluid} allows an unambiguous separation of equilibrium-superfluid and quasiparticle-relaxation effects, as long as data are taken over a wide frequency range.  The frequency-dependent superfluid density of \cecoin\ is plotted in Fig.~\ref{fig4}a, for frequencies ranging from 0.13 to 19.6~GHz.  We see that the frequency dependence of $1/\delta^2$ is indeed very strong and note the difficulty of isolating $1/\lambda^2_\mathrm{L}(T)$ from a measurement at any single microwave frequency.  $1/\lambda^2_\mathrm{L}(T)$ has a strong temperature dependence across the whole temperature range, in contrast to an $s$-wave superconductor\cite{Tinkham:1975un}, but is not strictly linear at low temperatures, as reported in previous studies\cite{Ormeno:2002p404,Chia:2003et,Ozcan:2003p400}.  This can be seen clearly in Fig.~\ref{fig4}b, where we show that $\Delta(1/\lambda^2_\mathrm{L}(T))$ is well described by a $T^{1.25}$ power law.  Note that our experiment directly measures the absolute penetration depth, and therefore the curvature is not the result of uncertainties in the absolute value of $\lambda_\mathrm{L}(T \to 0)$.

To resolve this puzzle, we allow for the possibility that the instantaneous diamagnetic response of \cecoin\ is temperature dependent, as discussed in the Introduction.  
In analogy with the superfluid density $1/\lambda_\mathrm{L}^2$, which is obtained from the static limit of $\omega \mu_0 \sigma_2$, we define a diamagnetic contribution $1/\lambda^2_\mathrm{d}$.   $1/\lambda^2_\mathrm{d}$ is proportional to the conduction electron density, $n$, and a Fermi surface average of the inverse of the effective mass, $m^\ast$, and can be accessed experimentally as the high-frequency limit of $\omega \mu_0 \sigma_2$:
\begin{equation}
\frac{1}{\lambda^2_\mathrm{d}}  = \mu_0 n e^2\left\langle\frac{1}{m^\ast}\right\rangle_\mathrm{FS} = \lim_{\omega \gg 1/\tau} \omega \mu_0 \sigma_2(\omega)\;.
\end{equation}

In our experiment, the condition that frequency be much larger than the quasiparticle relaxation rate, $1/\tau$, is satisfied at the lowest temperatures and highest frequencies.  In Fig.~\ref{fig4}a we see that, instead of becoming temperature independent, as expected for a conventional metal, the temperature slope of $\omega \mu_0 \sigma_2(\omega,T)$ changes sign at high frequencies, a clear indication that $1/\lambda^2_\mathrm{d}$ has temperature dependence in \cecoin.  The observed behaviour corresponds to an effective mass that increases on cooling, which was raised as a possibility in earlier work\cite{Ozcan:2003p400}.  This is likely a consequence of the proximity of \cecoin\ to quantum criticality\cite{Sidorov:2002cr}. 

Interestingly, de Haas--van Alphen measurements made at low fields (6--7~T) on \cecoin\ show an extreme departure from the standard Fermi-liquid, Lifshitz--Kosevich model, but are well described by a non-Fermi-liquid theory based on antiferromagnetic quantum criticality\cite{Mccollam:2005p396}.  Such behaviour can also be interpreted as a temperature-dependent quasiparticle mass.  At higher fields (13--15~T), where the material is tuned away from quantum criticality, the quantum oscillations revert to the standard Lifshitz--Kosevich form, in which quasiparticle mass is temperature independent.

The paramagnetic contribution to the superfluid density, $1/\lambda_\mathrm{p}^2$, which is sensitive to the nodal structure of the order parameter, can now be isolated via the relation
\begin{equation}
\frac{1}{\lambda^2_\mathrm{L}} = \frac{1}{\lambda^2_\mathrm{d}} - \frac{1}{\lambda^2_\mathrm{p}}\;.
\end{equation}
In Fig.~\ref{fig4}b we plot $1/\lambda_\mathrm{p}^2(T)$, using measurements of $1/\delta^2(\omega,T)$  at 19.6~GHz as a proxy for $1/\lambda^2_\mathrm{d}$ below 0.6~K.  The paramagnetic back-flow term, when properly isolated, is linear in temperature, providing direct evidence that the low energy quasiparticles have a nodal spectrum and giving strong support for a $d$-wave pairing state.  \\

\noindent {\bf Microwave spectroscopy.} At finite frequencies, the real part of the microwave conductivity, $\sigma_1$,  is due entirely to quasiparticle relaxation\cite{Bonn:2007hl}.
Data for $\sigma_1$ are presented in Fig.~\ref{fig5}a as a function of temperature. At all frequencies measured, $\sigma_1(T)$ shows an initial rise on cooling through $T_\mathrm{c} = 2.25$~K and a broad peak at intermediate temperatures.  Note that the form of this peak is very different from the conductivity coherence peak in an \mbox{$s$-wave} superconductor\cite{Tinkham:1975un}.  Theoretical curves\cite{MATTIS:1958p326} for the conductivity of an $s$-wave superconductor are shown in the inset of  Fig.~\ref{fig5}b: this behaviour has been confirmed by a number of classic experiments on conventional superconductors such as Al\cite{BIONDI:1959p338,Steinberg:2008p1904} and Pb\cite{Holczer:1991im}.  Instead of arising from BCS coherence factors, the conductivity peak in \cecoin\ is the result of a sharp collapse in quasiparticle scattering below \tc\ that  outpaces the gradual condensation of quasiparticles into the superfluid.  For comparison, we plot the $b$-axis conductivity of ultra-pure YBa$_2$Cu$_3$O$_{6.993}$\cite{Harris:2006p388} in Fig.~\ref{fig5}b.  The qualitative similarities of $\sigma_1(\omega,T)$ in the two materials are striking, revealing a deep connection in the underlying charge dynamics.  
However, the fact that the frequency scales are similar is somewhat puzzling, considering the large difference in energy scales such as \tc: the reason for this is that while inelastic scattering rates scale as \tc\ in the two materials, elastic scattering rates are determined by disorder and do not follow the same scaling.
Nevertheless, the similarities in $\sigma_1(\omega,T)$ provide strong support to the conjecture that these materials are different manifestations of the same correlated electron problem: two-dimensional $d$-wave superconductivity proximate to antiferromagnetism\cite{Scalapino:1986ec,Scalapino:1995p741,Monthoux:1999uj,Petrovic:2001p391}. 

Quantitative insights into the quasiparticle dynamics are obtained from conductivity frequency spectra, plotted in Fig.~\ref{fig6}.  The collapse in scattering inferred from $\sigma_1(T)$ can now be seen directly as the narrowing of $\sigma(\omega)$ on cooling.  We simultaneously fit to the real and imaginary parts of $\sigma(\omega)$ using a multi-component conductivity model, which builds on Eq.~\ref{Eq_complex_conductivity} by introducing a particular form for the quasiparticle conductivity spectrum, as described in Methods.
The first piece of this conductivity model is the superfluid term.  Its weight is proportional to $1/\lambda_\mathrm{L}^2(T)$, which is plotted in Fig.~\ref{fig4}a and is tightly constrained by the 0.13~GHz data.  For $\sigma_\mathrm{qp}(\omega)$ we use a two-component model consisting of a narrow, Drude-like spectrum, whose width gives the average quasiparticle relaxation rate $1/\tau$, and a frequency-independent background conductivity, $\sigma_\mathrm{bgnd}$.  The Drude-like spectrum has been modified by the inclusion of a conductivity frequency exponent, $y(T)$, which controls the detailed shape of the spectrum and allows for a distribution of quasiparticle relaxation rates\cite{Turner:2003p331,Harris:2006p388}.
The need for a two-component quasiparticle spectrum is most apparent in the low temperature traces in Fig.~\ref{fig6}a, which reveal narrow spectra, 3 to 4~GHz wide, riding on top of a broad background: a single-component spectrum cannot simultaneously capture the narrow, long-lived part of $\sigma_1(\omega)$ and the frequency-independent behaviour above 12~GHz.  In the absence of higher frequency data, we simply approximate the broad part of the spectrum as a constant.  We will later see that such a model is well motivated by the multi-band nature of \cecoin\cite{Tanatar:2005p405}.  Two of the fit parameters are plotted in Fig.~\ref{fig5}a: $\sigma_\mathrm{dc}(T)$, the zero-frequency limit of the quasiparticle conductivity; and the background conductivity $\sigma_\mathrm{bgnd}(T)$. The quasiparticle relaxation rate, $1/\tau(T)$, is plotted in Fig.~\ref{fig6}d. The conductivity frequency exponent, $y(T)$, is plotted in the inset of Fig.~\ref{fig6}d.\\

\noindent \textsf{\textbf{Discussion}}\\

The quasiparticle scattering dynamics of a superconductor separate broadly into two regimes: low temperature elastic scattering due to static disorder; and inelastic scattering, which becomes important at higher temperatures.  In a \dwave\ superconductor, both regimes are expected to carry signatures of the nodal quasiparticle spectrum, namely the linear energy dependence of the density of states\cite{Quinlan:1994vj,Walker:2000ux,Duffy:2001dg,Dahm:2005bv}.  

Quasiparticle scattering by impurities (elastic scattering) has been studied extensively in \dwave\ superconductors\cite{HIRSCHFELD:1993tf,HIRSCHFELD:1994p570,Walker:2000ux,Schachinger:2003p635}.  In the strong-scattering regime, impurities have a pair-breaking effect, causing a crossover to $T^2$ behaviour in the low temperature superfluid density.  We are able to rule out this type of scattering in \cecoin\ on the the basis of the data in Fig.~\ref{fig4}a.  In general, the quasiparticle scattering rate is determined by the phase space for recoil --- in the weak-scattering (Born) limit, the elastic scattering rate acquires a linear energy dependence (and therefore a linear temperature dependence) that reflects the structure of the clean \dwave\ density of states.  

Inelastic scattering can occur by a number of mechanisms, but the proximity to antiferromagnetism in \cecoin\ makes spin-fluctuation scattering an important candidate.  Curiously, both spin-fluctuation scattering and direct quasiparticle--quasiparticle scattering are expected to give rise to a $T^3$ temperature dependence in a \dwave\ superconductor\cite{Quinlan:1994vj,Walker:2000ux,Duffy:2001dg,Dahm:2005bv}.  On closer inspection, this simply reflects the fact that a spin fluctuation is a correlated electron--hole pair: in the superconducting state, correlations between electrons and holes weaken, and the spin-fluctuations increasingly resemble a dilute quasiparticle gas\cite{Walker:2000ux}. 
However, the $T^3$ scattering rate should not ordinarily be observable directly in electrical transport: Walker and Smith\cite{Walker:2000ux} have noted that charge currents require \emph{umklapp} processes to relax, in order that net momentum be removed from the electron system during scattering.  For $d$-wave quasiparticles, which at low temperatures are confined to the vicinity of the nodal points, momentum conservation leads to a minimum energy threshold, or ``\emph{umklapp} gap'', $\Delta_\mathrm{U}$, as illustrated in Fig.~\ref{fig7}.   Below this threshold, \emph{umklapp} processes cannot be excited and hence are frozen out, with $1/\tau_\mathrm{umklapp} \sim T^2 \exp(-\Delta_\mathrm{U}/k_\mathrm{B} T)$ at low temperatures.

We now turn to the data on \cecoin.  Immediately above \tc, $1/2 \pi \tau_\mathrm{n} \approx 120$~GHz.   (Normal-state quasiparticle lifetime, $\tau_\mathrm{n} = \sigma_1 \mu_0 \lambda_0^2$, is obtained by using the zero-temperature penetration depth as a gauge of plasma frequency.)  In temperature units, $\hbar/k_\mathrm{B} \tau_\mathrm{n} = 6$~K, several times larger than \tc, placing \cecoin\ in a similar regime of strong inelastic scattering as the cuprates\cite{VARMA:1989p468}.  On cooling into the superconducting state, $1/\tau(T)$ quickly drops into the low microwave range, where we can resolve it directly in the width of the conductivity spectra.  Below 1~K, in the disorder-dominated elastic regime, the relaxation rate reaches a residual value of $1/2 \pi \tau_0 = 3$~GHz.  The observation of a roughly temperature-independent relaxation rate implies an energy-independent phase space for recoil, and is difficult to understand in the context of simple \dwave\ superconductivity. 

 However, \cecoin\ displays prominent multiband effects\cite{Settai:2001p770}.  Parts of its Fermi surface have 2D character, with approximately cylindrical geometry (shown in blue in Fig.~\ref{fig7}d).  The Fermi surface also contains small, approximately isotropic Fermi pockets, with 3D character (shown in red in Fig.~\ref{fig7}d).  Quantum oscillation measurements reveal quite different masses for the 2D and 3D Fermi sheets, with the mass of the 3D pockets only weakly enhanced.  The coexistence of heavy and light-mass electron systems has been used in Ref.~\onlinecite{Tanatar:2005p405} to provide a simultaneous explanation of measurements of heat capacity ($\propto m^\ast$) and thermal conductivity ($\propto 1/m^\ast$). In the microwave measurements, the 3D Fermi pockets provide additional phase space for the scattering processes, as well as a natural explanation for the broad background conductivity, $\sigma_\mathrm{bgnd}$, observed in the $\sigma_1(\omega)$ spectra.  A distribution of relaxation rates naturally arises when there is a  strong variation of effective mass over the Fermi surface, as occurs in \cecoin\cite{Settai:2001p770,Hall:2001ua}; narrow conductivity spectra correspond to heavy quasiparticles and broad spectra to light ones\cite{Prange:1964gr}.  At the very lowest temperatures, there is a downturn in $1/\tau(T)$ that, while small, appears to be statistically significant.  This is consistent with the observation of temperature dependence of the quasiparticle effective mass.

In the intermediate temperature range, the relaxation rate is strongly temperature dependent and, as shown in Fig.~\ref{fig6}d, is well described by a sum of a temperature-independent elastic term and a $T^3$ inelastic term.  To facilitate a detailed comparison with spin-fluctuation theory, the functional form we fit to the relaxation rate is 
\begin{equation}
\frac{\hbar}{\tau(T)} = \frac{\hbar}{\tau_0} + A \frac{k_\mathrm{B}^3T^3}{\Delta^2(T)}\;,
\end{equation}
where the gap \mbox{$\Delta(T) = \Delta_0 \tanh\big(2.4 \sqrt{T_\mathrm{c}/T -1}\big)$} and $\Delta_0 = 3 k_\mathrm{B} T_\mathrm{c}$.  The prefactor $A$ is expected to be of order one\cite{Quinlan:1994vj}: numerical spin-fluctuation calculations obtain $A = 2.4$ for parameters relevant to optimally doped cuprates\cite{Dahm:2005bv}; fits to our \cecoin\ data give $A = 3.36$.  With these values so close, the charge dynamics of \cecoin\ appear to be consistent with a spin-fluctuation mechanism.

As noted above, inelastic contributions to electrical relaxation rate in a \dwave\ superconductor are expected to be suppressed by an \emph{umklapp} gap.  The observation of $T^3$ behaviour in \cecoin\ is therefore somewhat surprising, as it implies that the \emph{umklapp} gap is small.  As shown in Fig.~\ref{fig7}c, the \emph{umklapp} gap in a simple \dwave\ superconductor is determined by the location of the gap nodes with respect to the reciprocal lattice vectors\cite{Walker:2000ux}.  In a multi-band superconductor, in which superconductivity is weak on parts of the Fermi surface, the situation is more complex.  This is illustrated for the \cecoin\ Fermi surface in Fig.~\ref{fig7}d, in which we show how inter-band scattering can reduce the threshold for \emph{umklapp} processes.

Additional insights into the quasiparticle charge dynamics come from the conductivity frequency exponent, $y(T)$, which is used in our conductivity model to capture energy- or momentum-dependent scattering.  In the Drude limit ($y = 2$) all quasiparticles relax at the same rate.  Previous studies\cite{Turner:2003p331,Harris:2006p388} have shown that $y < 2 $ works well in capturing the phenomenology of $d$-wave superconductors, in which the quasiparticle relaxation rate has a strong energy dependence due to the Dirac-cone structure of the $d$-wave quasiparticle spectrum\cite{HIRSCHFELD:1993tf,HIRSCHFELD:1994p570,Schachinger:2003p635}.  Below 1.5~K, the best-fit value of $y(T)$ sits at the Drude limit, $y = 2$, implying that the low energy quasiparticles relax at approximately the same rate --- this is consistent with the weak temperature dependence of $1/\tau$ in this range and reflects the additional phase space for recoil provided by the multiband Fermi surface.  Above 1.5~K,  $y(T)$ drops quickly, falling below $1.2$ on the approach to \tc.  This indicates a rapidly broadening distribution of quasiparticle relaxation rates at higher energies, possibly associated with the development of hot spots on the Fermi surface due to spin-fluctuation scattering\cite{Rosch:1999ke}.  

To summarize the charge dynamics of \cecoin, our data fit well with a picture of heavy quasiparticles coexisting with uncondensed light quasiparticles\cite{Tanatar:2005p405}. The heavy quasiparticles experience a large decrease in scattering below \tc, and participate strongly in forming the superfluid, with only  7\% spectral weight remaining uncondensed as $T \to 0$.  The light quasiparticles undergo a much smaller decrease in scattering and have significant residual conductivity at low temperatures.  This suggests that the spectrum of fluctuations responsible for inelastic scattering, mass enhancement, and superconducting pairing couples strongly to the heavy parts of the Fermi surface, but much less efficiently to the light band.   
 
 Our results provide a new window into the low energy charge dynamics of \cecoin\ and uncover a complex interplay between $d$-wave superconductivity, multiband physics and quantum criticality.  The phenomena revealed can only be understood using measurements over a wide frequency range.  Many of the features are strongly reminiscent of the cuprates, confirming a close connection between these two classes of material.  An important difference is the observation of temperature-dependent quasiparticle mass, which not only resolves the issue of anomalous power laws in London penetration depth,  but shows that quantum criticality is not completely circumvented by the onset of superconductivity\cite{Laughlin:2001wa}.\\


\noindent \textsf{\textbf{Methods}}\\
\footnotesize
\noindent \textsf{\textbf{Surface impedance.}} Phase-sensitive measurements of microwave surface impedance, $Z_\mathrm{s} = R_\mathrm{s} + \I X_\mathrm{s}$, were made using resonator perturbation techniques\cite{1946Natur.158..234P,Altshuler:1963wq,Klein:1993p1129,Donovan:1993p1114,DRESSEL:1993p341,Huttema:2006p344,Bonn:2007hl}, with temperature-dependent changes in $Z_\mathrm{s}$ obtained from resonator frequency, $f_0$, and resonant bandwidth, $f_\mathrm{B}$, using the cavity perturbation approximation \mbox{$\Delta Z_\mathrm{s} = \Gamma \left( \Delta f_\mathrm{B}(T)/2 - \I \Delta f_0(T)\right)$}. Here $\Gamma$ is a resonator constant determined empirically from the known DC resistivity of \cecoin.\cite{Bauer:2006cr} At the lowest frequency, 0.13~GHz, surface reactance was measured using a tunnel-diode oscillator and was previously published in Ref.~\onlinecite{Ozcan:2003p400}.  At all other frequencies, surface impedance was measured using dielectric-resonator techniques, in a dilution-refrigerator-based variant of the apparatus described in Ref.~\onlinecite{Huttema:2006p344}.  The absolute surface resistance was obtained at each frequency using an \emph{in-situ} bolometric technique, by detecting the synchronous rise in temperature when the sample was subjected to a microwave field of known, time-varying intensity\cite{Turner:2004p332}.  Thermal expansion effects make a small contribution to the apparent surface reactance, $\Delta X_\mathrm{s}^\mathrm{th} \approx -\frac{\omega \mu_0 c}{2}\beta(T)$, where $c$ is the thickness of the sample and $\beta(T)$ is the volume coefficient of thermal expansion.  This was corrected for using thermal expansion data from Ref.~\onlinecite{Takeuchi:2002wf}.\\

\noindent \textsf{\textbf{Absolute surface reactance.}} The absolute surface reactance was obtained at 2.91~GHz by matching $R_\mathrm{s}(T)$ and $X_\mathrm{s}(T)$ between 10~K and 35~K, a temperature range in which the imaginary part of the normal-state conductivity is small and $R_\mathrm{s} = X_\mathrm{s} = \sqrt{\omega \mu_0 \rho_\mathrm{dc}/2}$. To a first approximation, $X_\mathrm{s} \approx \omega \mu_0 \lambda_\mathrm{L}$ is used to obtain the surface reactance at the other frequencies.  This estimate is refined by taking into account the quasiparticle contribution to $X_\mathrm{s}$.  We carry this procedure out at $T = 0.1$~K, using the following self-consistent method.  The quasiparticle contribution to $\sigma_2$ is initially set to zero, so that $\sigma_2(\omega)$ arises purely from the superfluid conductivity, $\sigma_\mathrm{s} = 1/\I \omega \mu_0 \lambda_\mathrm{L}^2$, with $\lambda_\mathrm{L}$ obtained from the 2.91~GHz $X_\mathrm{s}$ data. The local electrodynamic relation, $Z_\mathrm{s} =\sqrt{\I \omega \mu_0/\sigma }$, is used to obtain $X_\mathrm{s}(\omega)$ from the measured $R_\mathrm{s}(\omega)$ and the calculated $\sigma_2(\omega)$. This step is carried out without any explicit knowledge of $\sigma_1(\omega)$. From $X_\mathrm{s}(\omega)$ and $R_\mathrm{s}(\omega)$ we obtain $\sigma_1(\omega)$, again using the local electrodynamic relation. A Drude-like spectrum, $\sigma_1^\mathrm{qp}(\omega) = \sigma_0/(1 + \omega^2 \tau^2) + \sigma_\mathrm{bgnd}$, is fit to $\sigma_1(\omega)$. The corresponding imaginary part, $\sigma_2^\mathrm{qp}(\omega) = \sigma_0 \omega \tau/(1 + \omega^2 \tau^2)$, provides an estimate of the quasiparticle contribution to $\sigma_2(\omega)$. The total imaginary conductivity is the sum of $\sigma_2^\mathrm{qp}(\omega)$, and a superfluid term of the same form as in step~1, but with $\lambda_\mathrm{L}$ adjusted to make $\sigma_2(\omega)$ consistent with the directly determined value of $X_\mathrm{s}$ at 2.91~GHz. The refined estimate of $\sigma_2(\omega)$ is inserted into the beginning of the procedure, and the process is iterated to self consistency.\\

\noindent \textsf{\textbf{Microwave conductivity.}} The complex microwave conductivity, $\sigma_1(\omega) - \I \sigma_2(\omega)$, is obtained from the surface impedance using the local electrodynamic relation. At each temperature, the real and imaginary parts of $\sigma$ are simultaneously fit to a three-component model consisting of a superfluid term; a broad background conductivity; and a narrow, Drude-like spectrum:
\begin{equation}
\sigma(\omega) = \frac{1}{\I \omega \mu_0 \lambda^2_\mathrm{L}} + \sigma_\mathrm{bgnd} + \left(\frac{\sigma_0}{1 + (\omega \tau)^y} - \I \sigma_\mathrm{KK}(\omega;\sigma_0, \tau, y)\right).
\label{Eq_conductivity}
\end{equation}
The parameters of the model are: the superfluid density, $1/\lambda_\mathrm{L}^2(T)$; the background conductivity, $\sigma_\mathrm{bgnd}(T)$; the magnitude of the Drude-like spectrum, $\sigma_0(T)$; the relaxation time, $\tau(T)$; and the conductivity exponent, $y(T)$, constrained to the interval $1 < y \le 2$.  $\sigma_\mathrm{KK}(\omega)$ denotes the imaginary part of the Drude-like spectrum, obtained using a Kramers--Kr\"onig transform.\\

\noindent \textsf{\textbf{Samples.}}  High quality single crystals of \cecoin\ were grown by a self-flux method in excess In\cite{Petrovic:2001p391,Kim:2001gd}.  The microwave measurements were carried out on a mm-sized platelet with naturally formed, mirror-like $a$--$b$ plane faces.  This sample was the same as that used in Ref.~\onlinecite{Ozcan:2003p400}.  The width of the (003) x-ray rocking curve was 0.014$^\circ$, indicating high crystallinity.  Electron-probe microanalysis gives an average composition of Ce$_{1.02(1)}$Co$_{0.99(1)}$In$_{4.99(1)}$, homogeneously throughout the bulk of the crystal, indicating that the samples are single-phase and highly stoichometric.

\vspace{3 mm}
\normalsize
\noindent \textsf{\textbf{Acknowlegdements}}\\
\footnotesize
We thank M.~Dressel, S.~R.~Julian and M.~Scheffler for discussions and correspondence. Research support for the experiments was provided by the Natural Science and Engineering Research Council of Canada and the Canadian Foundation for Innovation.  Research support for sample preparation was provided by the Division of Materials Science and Engineering of the U.S. Department of Energy Office of Basic Energy Sciences.\\

\normalsize
\noindent \textsf{\textbf{Author contributions}}\\
\footnotesize
P.J.T., W.A.H., C.J.S.T., S.\"O., P.R.C., E.T., N.C.M., K.J.M., A.J.K.\ and D.M.B.\ designed and set up the dilution-refrigerator-based systems for microwave spectroscopy.  C.J.S.T., W.A.H., P.J.T., S.\"O., N.C.M.\ and D.M.B.\  carried out the experiments.  C.J.S.T.\ carried out the data analysis.  J.L.S.\ prepared the sample of CeCoIn$_5$.  D.M.B.\ wrote the paper and supervised the project.\\ 

\normalsize
\noindent \textsf{\textbf{Additional information}}\\
\footnotesize
{\bf Competing financial interests.} The authors declare no competing financial interests.\\

\normalsize
\noindent \textsf{\textbf{Corresponding author}}\\
\footnotesize
Correspondence and requests for materials should be addressed to D.M.B. (email: dbroun@sfu.ca).\\

\clearpage

\normalsize

\end{document}